\documentclass[conference]{IEEEtran}
\IEEEoverridecommandlockouts
\usepackage{cite}
\usepackage{amsmath,amssymb,amsfonts}
\usepackage{graphicx}
\usepackage{textcomp}
\usepackage{xcolor}
\usepackage{xspace}
\usepackage{algorithm}
\usepackage{algpseudocode}
\usepackage{tabularx}
\usepackage{svg}
\usepackage{subcaption}
\usepackage{enumitem}
\usepackage{multirow}
\usepackage{booktabs}
\usepackage{array}
\usepackage{stfloats}
\usepackage{float}
\usepackage[section]{placeins}
\usepackage{amsthm}
\usepackage{adjustbox}
\usepackage{comment}
\usepackage[T1]{fontenc}
\usepackage{soul}
\sethlcolor{yellow}
\def\BibTeX{{\rm B\kern-.05em{\sc i\kern-.025em b}\kern-.08em
    T\kern-.1667em\lower.7ex\hbox{E}\kern-.125emX}}

\newcommand{\cmark}[1]{\textcircled{#1}}

\newcommand{\zn}[1]{}
\newcommand{\tk}[1]{}
\newcommand{\note}[1]{}
\newcommand{\info}[1]{}

\newcommand{\name}{GreenGNN\xspace}

\begin{document}
\title{\name: Energy-Aware Windowed Communication Optimization for Distributed GNN Training}

\author{
\IEEEauthorblockN{Arefin Niam}
\IEEEauthorblockA{
Tennessee Technological University\\
Cookeville, Tennessee, USA\\
aniam42@tntech.edu
}
\and
\IEEEauthorblockN{Tevfik Kosar}
\IEEEauthorblockA{
University at Buffalo\\
Buffalo, New York, USA\\
tkosar@buffalo.edu
}
\and
\IEEEauthorblockN{M. S. Q. Zulkar Nine}
\IEEEauthorblockA{
Tennessee Technological University\\
Cookeville, Tennessee, USA\\
mnine@tntech.edu
}
}

\maketitle
\begin{abstract}
Large-scale graph neural network (GNN) training often requires distributed clusters because graph structure and feature tensors no longer fit in a single node's memory. In sampling-based training, each mini-batch expands into a receptive field that spans partitions and triggers thousands of remote feature fetches per epoch. This wastes energy for two main reasons: each small RPC pays a fixed initiation and protocol cost, and GPUs continue drawing substantial baseline power while waiting for remote features. We present \name, an energy-aware distributed GNN training system that reduces communication energy by exploiting the bursty, short-lived temporal locality of neighbor sampling. \name groups training into windows of $W$ consecutive mini-batches, stages each window's hot features in a local cache, and merges remote requests from each partition owner into a small number of bulk transfers. This amortizes RPC overhead across many features while preserving an on-demand path for cache misses. Because window size controls the trade-off between communication amortization and hot-set staleness, \name selects $W$ offline using a discrete-event simulator that replays a deterministic one-epoch access trace with a hybrid energy model. We implement \name on DGL and evaluate it on a 4-node GPU cluster with benchmark datasets. Across datasets and batch sizes, \name reduces total system energy by 27--43\% relative to baseline while improving end-to-end throughput by up to 3.9$\times$. GPU energy drops by 36--71\%, driven by fewer RPC initiations and lower GPU stall time.
\end{abstract}


\section{Introduction}
\label{sec:intro}

Deep learning has achieved remarkable success on Euclidean data, but many important real-world workloads are fundamentally graph-structured. Modern recommendation systems~\cite{ying2018graph,fan2019graph}, financial fraud detection pipelines~\cite{dou2020enhancing}, and biomedical interaction networks~\cite{zhang2021bio} all operate over relational data. Graph Neural Networks (GNNs) have therefore emerged as the standard framework for learning on graphs~\cite{wu2020comprehensive,hamilton2017inductive}. In production, these graphs already contain hundreds of millions to billions of nodes and edges~\cite{ying2018graph,leskovec2016snap}, and industrial deployments have reported graphs with more than one trillion edges~\cite{ching2015one}. At this scale, graph structure and feature tensors no longer fit in the memory of a single accelerator, making distributed training across multiple machines necessary~\cite{besta2024parallel,zheng2020distdgl,gandhi2021p3}.

Distributed GNN training is challenging because its communication pattern is fundamentally irregular. Unlike dense DNNs that operate on contiguous tensors, GNNs repeatedly aggregate features from sampled neighbors along the graph topology~\cite{hamilton2017inductive}. In sampling-based distributed training, each mini-batch expands into a receptive field that often spans multiple partitions, triggering many remote feature lookups. This cross-partition traffic is a major bottleneck in distributed GNN training~\cite{zheng2020distdgl,gandhi2021p3,wan2022dgs,shao2024distributed}. Prior systems such as DistDGL~\cite{zheng2020distdgl}, P3~\cite{gandhi2021p3}, BGL~\cite{liu2023bgl}, and Legion~\cite{sun2023legion} have improved performance through pipelining, partition optimization, and feature caching. However, these systems are designed primarily to improve throughput. They pay far less attention to the energy cost of the communication pattern itself.

\begin{figure}[!ht]
  \vspace{-3mm}
  \centering
  \includegraphics[width=\columnwidth]{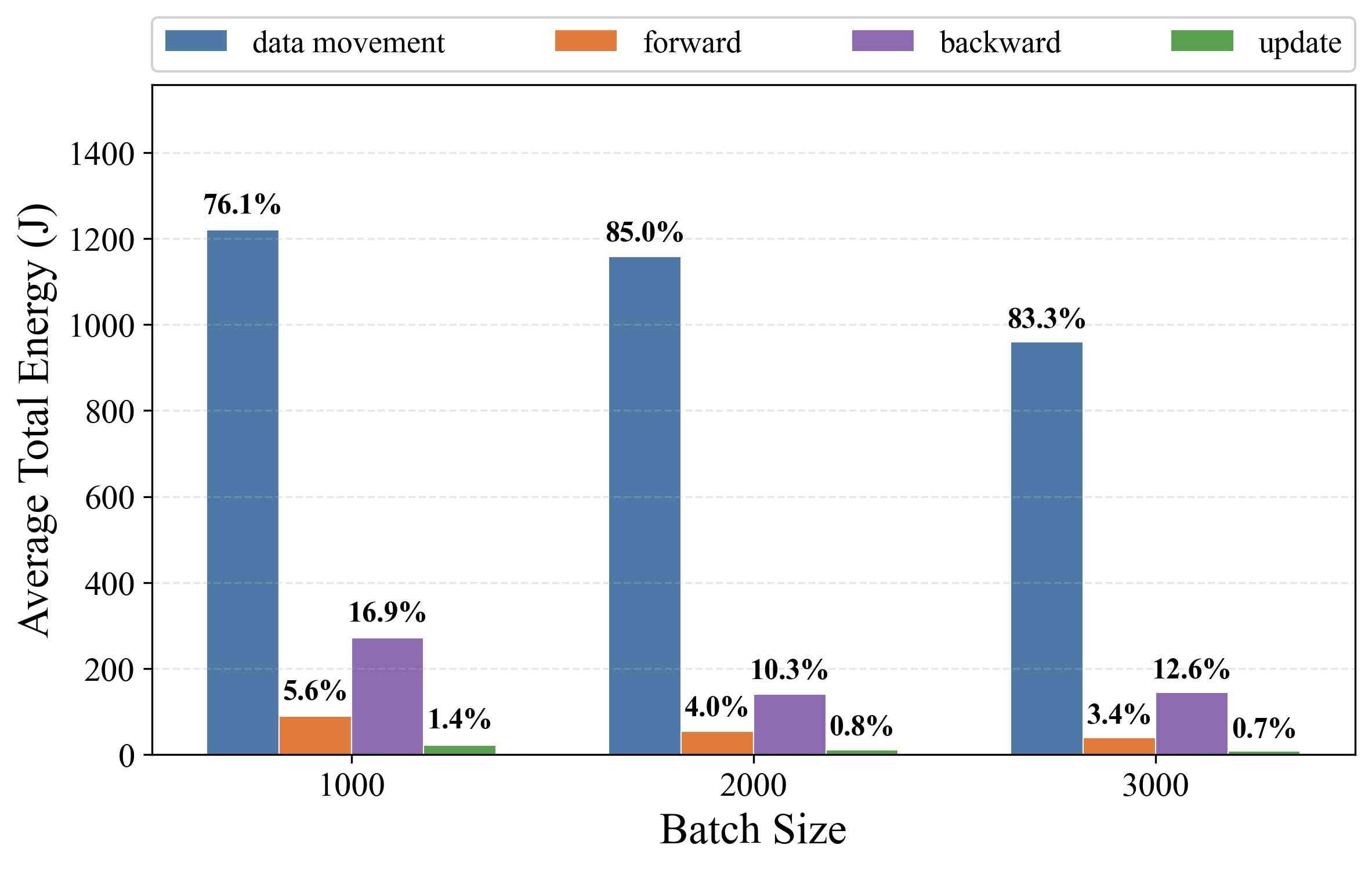}
  \caption{Energy breakdown of distributed GraphSAGE training on the OGBN-Products dataset.}
  \label{fig:energy_breakdown}
  \vspace{-2mm}
\end{figure}

This gap matters because communication dominates energy consumption in distributed GNN training. In our profiling of distributed GraphSAGE\cite{hamilton2017inductive}, we find that \textbf{data movement accounts for 76--85\% of total system energy} across batch sizes, while forward and backward computation together account for only 4--17\% (Figure~\ref{fig:energy_breakdown}). The inefficiency comes from two sources. First, each small RPC incurs a fixed initiation cost in CPU cycles, interrupt handling, and protocol processing, regardless of how little data it transfers. Second, GPUs continue to draw substantial baseline power while waiting for remote features to arrive. In other words, the system wastes energy both in issuing many fine-grained requests and in keeping accelerators idle during the resulting stalls.

Existing approaches reduce remote accesses, but they do not directly address this energy inefficiency. Partitioning methods such as min-cut~\cite{karypis1998fast} and replication strategies such as halo construction~\cite{zheng2020distdgl} reduce communication volume, but they do not change the fine granularity of many remote requests. Static feature caches~\cite{lin2020pagraph,yang2022gnnlab} place selected hot features in GPU memory with low runtime overhead, but their contents are fixed and become less effective as the hot set evolves during training on power-law graphs~\cite{backstrom2012four}. Dynamic caches~\cite{liu2023bgl} adapt to changing access patterns through replacement policies such as LRU, LFU, and FIFO, but they must track a large number of accesses and often synchronize metadata across workers~\cite{zhang2023two}. That overhead increases CPU activity and, in turn, energy consumption. Two-level cache designs~\cite{zhang2023two} increase effective cache capacity, but still rely on per-access tracking and optimize a sum-of-costs objective that does not reflect the slowest-partition bottleneck in distributed execution. Overall, prior work treats communication mainly as a latency and throughput problem. It does not treat energy as a first-class optimization objective, nor does it exploit the temporal structure of remote accesses in a way that is most beneficial for energy efficiency.

\name is built on a simple observation about distributed GNN training. Neighbor sampling repeatedly revisits dense graph regions and high-degree hubs, which means that a relatively small set of remote features is reused heavily over a short sequence of consecutive mini-batches before quickly falling out of the working set. We quantify this effect in Section~\ref{sec:results}. On Reddit dataset, caching the top 10\% most frequently accessed remote nodes within a window of $W{=}16$ mini-batches captures more than 85\% of all remote feature requests. This bursty and short-lived locality suggests a middle ground between static caching and fine-grained dynamic caching. Instead of updating the cache on every access, the system can refresh it periodically at window boundaries.

Based on this insight, \name adopts \emph{window-based caching}. It divides each epoch into windows of $W$ consecutive mini-batches. At the beginning of each window, it identifies the hot remote features for that window from a precomputed access trace, stages them into a local cache through a small number of bulk transfers from each partition owner, and serves most requests within the window directly from the cache. A lightweight on-demand path handles the remaining misses. This design transforms communication from thousands of small per-batch RPCs into a few large per-window transfers. As a result, it amortizes RPC initiation overhead across many features and reduces both CPU-side packet processing energy and GPU idle energy.

The key design question is how to choose the window size $W$. If $W$ is too small, bulk transfers do not provide enough amortization benefit. If $W$ is too large, the hot set becomes stale, cache misses increase, and GPU memory pressure rises. Searching for $W$ online is too expensive because each candidate requires a full training run. \name instead exploits the determinism of the access pattern under fixed seeds and partitioning. It collects a one-epoch access trace from a lightweight calibration run and replays that trace in a discrete-event simulator that estimates the energy cost of each candidate window size. The simulator uses a hybrid model that combines analytical energy estimation with a learned rank-correction component trained using pairwise ranking loss.

We implement \name on top of PyTorch and DistDGL and evaluate it on Reddit, OGBN-Products, and OGBN-Papers100M. Across datasets and batch sizes, \name reduces total system energy by 27--43\% relative to on-demand DistDGL, with GPU energy reductions of 36--71\%. We also find that the energy-window relationship is consistently convex, which confirms the expected trade-off between transfer amortization and hot-set staleness. The simulator-guided autotuner achieves Kendall's $\tau$ of 0.62--1.00 across all tested settings.

This paper makes the following contributions:
\begin{itemize}
  \item \textbf{Energy characterization of distributed GNN training.}
  We present a systematic energy breakdown of sampling-based distributed mini-batch GNN training and show that per-RPC initiation overhead and GPU stall power together account for more than 80\% of total system energy in our setting (Section~\ref{sec:motivation}).

  \item \textbf{Window-based caching with bulk transfer consolidation.}
  We introduce a caching design that exploits bursty temporal locality by grouping training into windows of $W$ mini-batches and replacing thousands of fine-grained RPCs with a small number of bulk transfers per partition owner, without per-access tracking or cross-worker metadata synchronization (Section~\ref{sec:design}).

  \item \textbf{A simulator-guided energy autotuner.}
  We develop an offline autotuner that combines a physics-based energy model with learning-to-rank calibration to select energy-efficient window sizes from a single-epoch access trace, achieving 6/9 top-1 accuracy and Kendall's $\tau$ up to 1.00 across datasets and batch sizes (Section~\ref{sec:results}).

  \item \textbf{Comprehensive evaluation.}
  We provide detailed CPU, GPU, and per-partition energy analysis on three datasets and show that \name reduces total system energy by up to 43\% and GPU energy by up to 71\% over baseline DistDGL while preserving model accuracy and matching or improving end-to-end throughput (Section~\ref{sec:results}).
\end{itemize}
\section{Motivation}
\label{sec:motivation}

Figure~\ref{fig:energy_breakdown} shows that data movement dominates the energy budget of distributed GNN training. In this section, we explain why this happens and connect the observed energy cost to two underlying physical mechanisms that appear consistently in our profiling results.

\subsection{Per-Request Initiation Overhead}
\label{subsec:init_overhead}

The energy cost of a remote feature fetch consists of two parts: a fixed \emph{initiation cost} $E_{\text{init}}$, which includes CPU interrupt handling, kernel crossings, and protocol processing, and a variable \emph{payload cost} $E_{\text{payload}}$, which grows with the number of features transferred. For the small transfers produced by GNN neighbor sampling, the fixed component is often the dominant one.

\begin{figure}[t]
  \centering
  \includegraphics[width=\columnwidth]{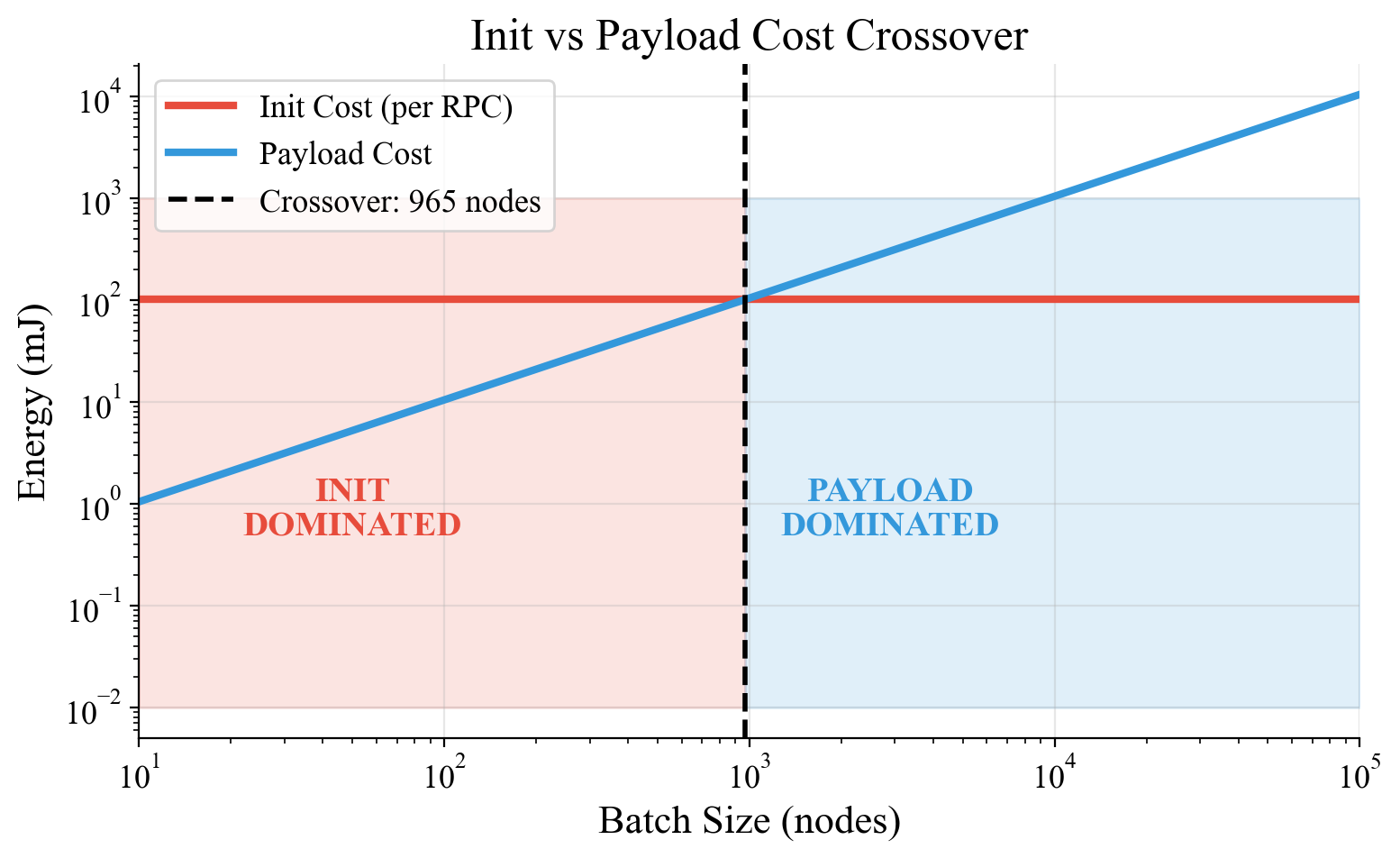}
  \caption{Initiation vs.\ payload energy as a function of transfer size. The dashed line marks the crossover ($\approx$965 nodes) between the initiation-dominated and payload-dominated regimes.}
  \label{fig:crossover}
\end{figure}

To quantify this effect, we profile RPC energy on our cluster across a range of payload sizes. At GNN-typical request sizes, which usually involve tens to low hundreds of remote nodes per RPC, initiation accounts for 49--99\% of total per-RPC energy. The crossover point at which payload cost begins to exceed initiation cost occurs at about 965 nodes, as shown in Figure~\ref{fig:crossover}. This threshold is much larger than the size of a typical per-batch remote request. As a result, existing systems operate mostly in the initiation-dominated regime and repeatedly pay a high fixed energy cost for moving relatively little data.

The consequence is straightforward. If many small requests are consolidated into a few larger transfers, the fixed cost can be amortized across thousands of features and the operating point moves into the payload-dominated regime, where energy scales more efficiently with transfer volume.

This argument extends beyond TCP. RDMA-capable interconnects reduce software overhead but still require per-operation work-request posting and completion through the NIC queue-pair interface; most distributed GNN frameworks, including DistDGL, rely on TCP-based RPC by default~\cite{zheng2020distdgl}. The core problem lies in the granularity of the access pattern, not only in the choice of transport.

\subsection{GPU Stall Energy}
\label{subsec:stall_energy}

Modern datacenter GPUs draw substantial baseline power even when they are not actively computing~\cite{nvidia2022h100}. In distributed GNN training, GPUs often wait for remote features before computation can proceed, and the baseline draw during these waiting periods accumulates into a significant energy cost.
This highlights an important distinction between throughput and energy optimization. Pipelining can overlap communication with computation and improve throughput, but does not reduce the total data fetched or the waiting cost of the communication pattern. Reducing stall energy requires reducing the number and duration of remote accesses, not simply masking them behind computation.

These two mechanisms, per-RPC initiation overhead and GPU stall
power, together account for the dominant share of system energy in
our measurements and motivate the design of \name. Reducing the
number of network round-trips is the highest-leverage path to
improving energy efficiency.
\section{Related Work}
\label{sec:related_work}

We organize prior work by how distributed GNN systems manage remote feature access, then highlight why these designs remain insufficient for energy-aware optimization.

\subsection{Partition and Communication Optimization}
 
A common approach to reducing cross-partition traffic is better graph
partitioning. DistDGL~\cite{zheng2020distdgl} uses METIS-based
edge-cut minimization~\cite{karypis1998fast} with 1-hop halo nodes,
DGCL~\cite{cai2021dgcl} improves the collective communication
layer, and GraNNDis~\cite{song2024granndis} introduces
expansion-aware sampling with one-hop graph masking. These systems
reduce the volume of remote traffic but do not change its fine
granularity: each mini-batch still generates many small RPCs, each
paying a fixed initiation cost. A separate line of work improves
throughput by overlapping communication with computation.
P3~\cite{gandhi2021p3} restructures feature and gradient movement
through a push-pull execution model, and
PipeGCN~\cite{wan2022pipegcn} uses stale features from a previous
iteration to decouple forward and backward passes. These techniques
reduce training time but do not reduce the number of RPC initiations
or the total data movement, so better overlap does not necessarily
translate into proportional energy savings
(Section~\ref{subsec:stall_energy}).

\subsection{Feature Caching}
\label{subsec:rw_caching}

Caching reduces redundant remote fetches by exploiting temporal locality.
Static designs such as PaGraph~\cite{lin2020pagraph} and GNNLab~\cite{yang2022gnnlab} use offline heuristics (node degree or pre-sampling frequency) to place hot features in GPU memory before training begins. These methods impose little runtime overhead but lose effectiveness on power-law graphs as the hot set shifts across seed batches.

Dynamic designs trade higher runtime cost for adaptability. BGL~\cite{liu2023bgl} evaluates LRU, LFU, and FIFO replacement, Legion~\cite{sun2023legion} adds NVLink-aware hierarchical caching, and XGNN~\cite{tang2024xgnn} unifies GPU and host memory through a global store abstraction. Zhang et al.~\cite{zhang2023two} expand capacity with a two-level GPU--CPU cache but optimize a sum-of-costs objective that ignores the slowest-partition bottleneck. All dynamic approaches incur per-access tracking and, in distributed settings, cross-worker metadata synchronization---overhead paid by the CPU that contributes directly to energy consumption, though none of these works quantify it from an energy perspective.

\subsection{The Gap}
\label{subsec:rw_gap}

\begin{table}[!htb]
\centering
\caption{Comparison of distributed GNN systems along key design axes.}
\label{tab:related_comparison}
\footnotesize
\begin{tabular}{lccccc}
\toprule
\textbf{System} & \textbf{Cache} & \textbf{Tracking} & \textbf{Bulk} & \textbf{Energy} \\
 & \textbf{Type} & \textbf{Granularity} & \textbf{Transfers} & \textbf{Aware} \\
\midrule
DistDGL~\cite{zheng2020distdgl}      & None     & ---        & No  & No \\
P3~\cite{gandhi2021p3}               & Limited  & ---        & No  & No \\
PaGraph~\cite{lin2020pagraph}        & Static   & None       & No  & No \\
GNNLab~\cite{yang2022gnnlab}         & Static   & None       & No  & No \\
BGL~\cite{liu2023bgl}                & Dynamic  & Per-access & No  & No \\
Legion~\cite{sun2023legion}          & Dynamic  & Per-access & No  & No \\
XGNN~\cite{tang2024xgnn}             & Dynamic  & Per-access & No  & No \\
Zhang et al.~\cite{zhang2023two}     & 2-Level  & Per-access & No  & No \\
RapidGNN~\cite{rapidgnn2025}         & Epoch    & Per-epoch  & Yes & No \\
\textbf{\name}                       & \textbf{Window} & \textbf{Per-window} & \textbf{Yes} & \textbf{Yes} \\
\bottomrule
\end{tabular}
\end{table}

Table~\ref{tab:related_comparison} summarizes the design space. Two limitations are clear. First, no existing distributed GNN system treats energy as a first-class optimization objective, despite the very different communication pattern GNN training exhibits compared to dense DNN collectives~\cite{shao2024distributed}. Second, prior systems do not exploit the temporal granularity at which remote access locality appears: the hot set is bursty and short-lived, calling for periodic cache refreshes at window boundaries rather than whole-run static placement or per-access tracking~\cite{rapidgnn2025}.

\name addresses both gaps. It builds on RapidGNN~\cite{rapidgnn2025}, which introduced deterministic sampling and trace-based prefetching for throughput. \name reorients the system around energy efficiency by adding a physics-based energy model with learned rank correction for offline window-size selection and consolidating per-batch RPCs into per-window bulk transfers.
\section{System Design}
\label{sec:design}

\begin{figure}[t]
  \centering
  \vspace{-12pt}
  \includegraphics[width=\columnwidth]{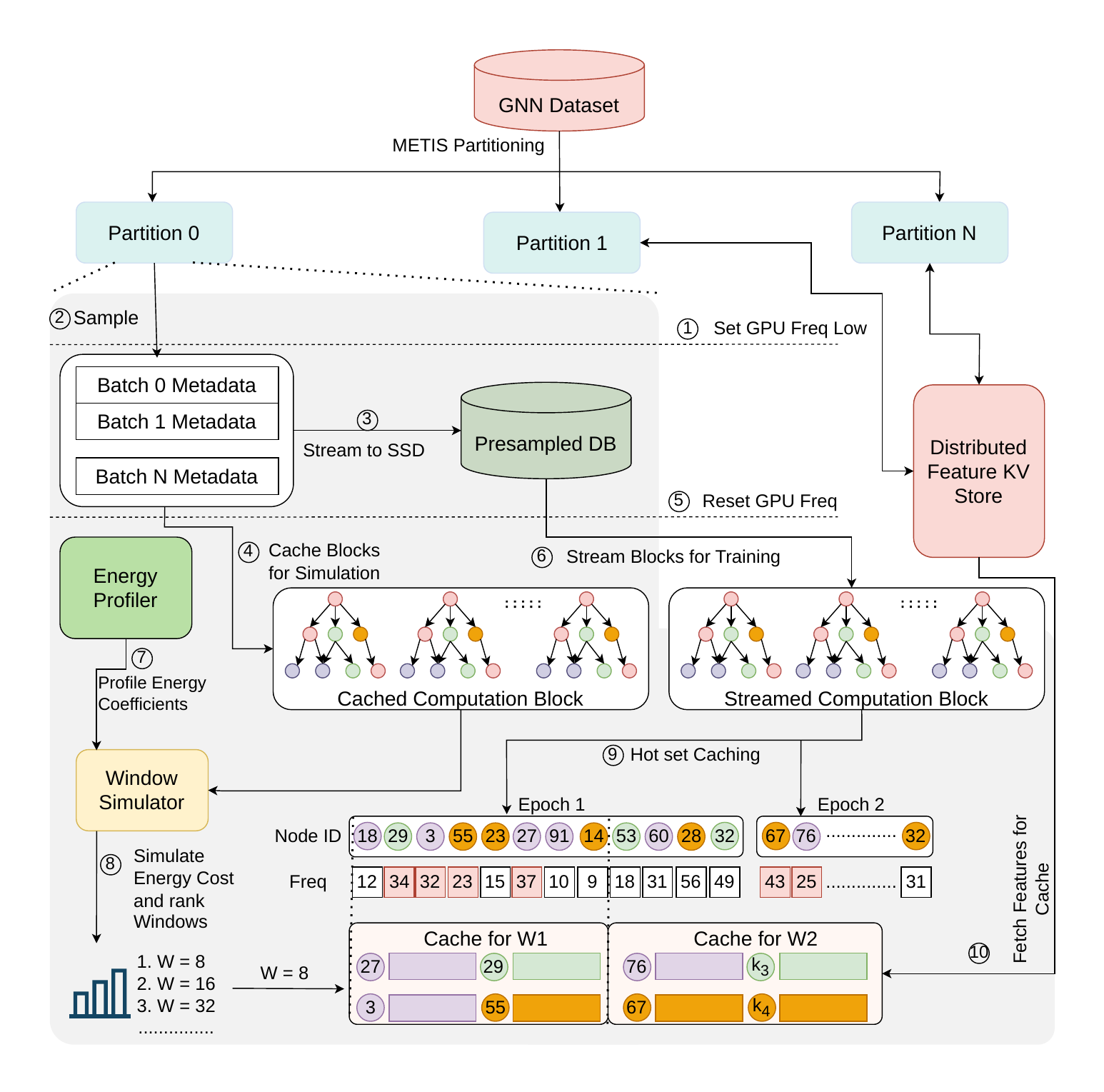}
  \caption{\name architecture overview. Circled numbers indicate execution order. Steps \textcircled{1}--\textcircled{8} constitute the offline phase (trace collection, energy profiling, and window selection); steps \textcircled{9}--\textcircled{10} execute at runtime (hot-set caching and feature fetching).}
  \label{fig:architecture}
\end{figure}

\name operates in two phases.
An \emph{offline} phase collects a deterministic access trace, profiles the cluster's energy characteristics, and selects an energy-optimal window size~$W^*$ through simulation.
A \emph{runtime} phase uses $W^*$ to drive pipelined window-based caching, in which a dedicated prefetcher builds the next window's cache in the background while the current window's mini-batches train on the GPU.
Figure~\ref{fig:architecture} shows the full pipeline.

\noindent\textbf{Notation.}
$\mathcal{O}$ is the set of remote partition owners; each remote node $u$ belongs to exactly one owner $m \in \mathcal{O}$.
Mini-batches are indexed $b = 1,\dots,B$, and $N_b$ is the set of nodes whose features batch~$b$ requires.

\subsection{Offline Phase}
\label{subsec:offline}

The offline phase has three jobs: collect the access trace, profile the hardware, and pick the best window size.

\vspace{0.3em}
\noindent\textbf{Trace collection (steps \textcircled{1}--\textcircled{5}).}
\name lowers the GPU clock (step~\textcircled{1}), runs one epoch of deterministic neighbor sampling (step~\textcircled{2}), and streams the resulting per-batch node-ID lists to a Presampled~DB on local SSD (step~\textcircled{3}).
Because seeds are fixed, this single-epoch trace $\mathcal{T} = \{(b, u) \mid u \in N_b\}$ fully determines the access pattern for all future epochs.
The Presampled~DB then feeds both the simulation path (step~\textcircled{4}) and, later, the runtime training path (step~\textcircled{6}).
The GPU clock is restored before training begins (step~\textcircled{5}).

\vspace{0.3em}
\noindent\textbf{Hardware profiling (step \textcircled{7}).}
The Energy Profiler runs microbenchmarks against each remote owner $m$ to measure an initiation energy $\epsilon_{\text{init}}(m)$ (the fixed cost of one RPC), a payload energy function $\epsilon_{\text{payload}}(m, x)$ (marginal cost of transferring $x$ features), corresponding latency surrogates $\lambda_{\text{init}}(m)$ and $\lambda_{\text{payload}}(m, x)$, and the aggregate bandwidth ceiling $\mathrm{BW}$.
Section~\ref{subsec:impl_profiler} details the profiling pipeline.

\vspace{0.3em}
\noindent\textbf{Simulation and window selection (step \textcircled{8}).}
A discrete-event simulator replays the trace for each candidate window size $W \in \mathcal{W}$ (e.g., $\{1, 2, 4, 8, 16, 32, 64\}$).
For each $W$, it partitions the epoch into $\lceil B/W \rceil$ windows, selects the $n_{\text{hot}}$ most frequently accessed remote nodes per window as the hot set $H_k$, and tallies two types of communication events:

\begin{itemize}
  \item \emph{Cache rebuilds}: at each window boundary, one bulk RPC per partition owner carrying the hot-set entries not already cached. For owner $m$, the rebuild delta at window $k$ is $\Delta_{k,m} = (H_k \cap N_m) \setminus (C_{k-1} \cap N_m)$, where $C_{k-1}$ is the previous cache state.
  \item \emph{Residual misses}: within a window, on-demand RPCs for nodes outside the hot set. For batch $b$ and owner $m$, the miss set is $M_{b,m} = (N_b \cap N_m) \setminus (H_k \cap N_m)$.
\end{itemize}

\begin{figure}[t]
  \centering
  \includegraphics[width=\columnwidth]{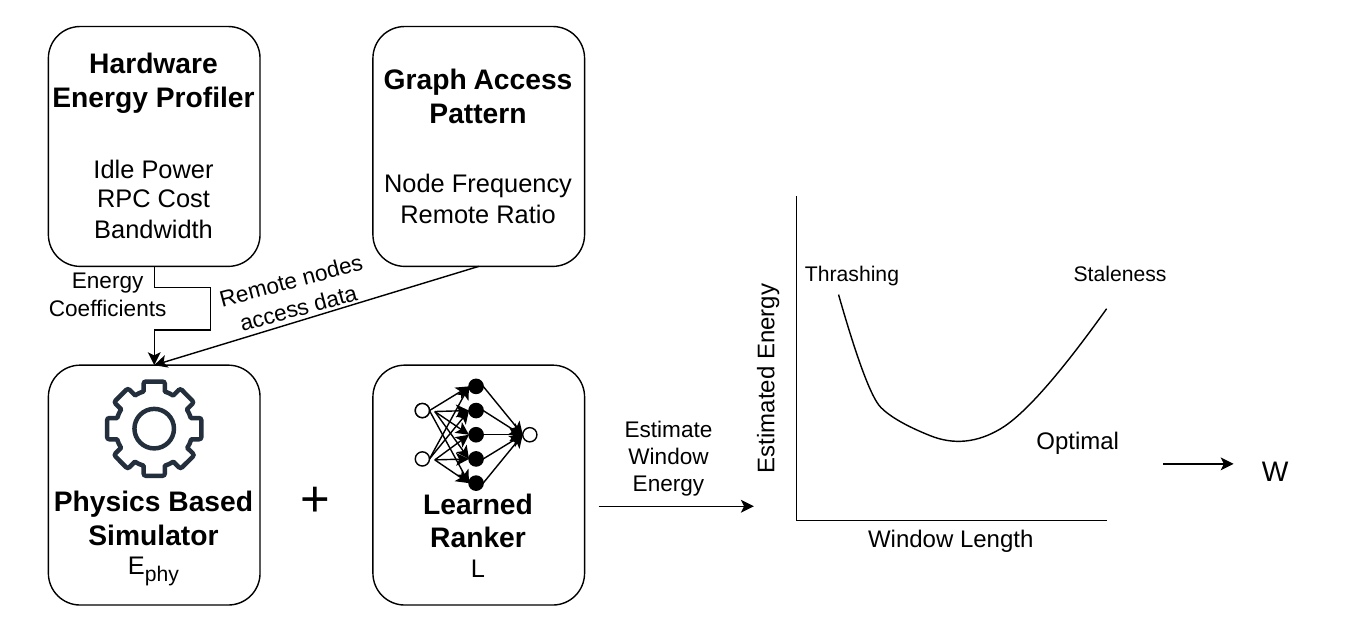}
  \caption{\name simulator workflow. For each candidate $W$, the simulator computes a physics-based estimate $E_{\text{phy}}(W)$ and applies a learned rank correction $L(W)$ to produce $E_{\text{net}}(W)$. The energy curve is convex: small windows incur frequent rebuilds, large windows suffer hot-set staleness, and $W^*$ lies in the valley.}
  \label{fig:simulator_workflow}
\end{figure}

These event counts feed a hybrid energy cost model described below.
Figure~\ref{fig:simulator_workflow} illustrates the simulator workflow.

\subsection{Hybrid Energy Cost Model}
\label{subsec:cost}

\name scores each candidate $W$ using a model that pairs an analytical energy estimate with a learned rank correction.

\vspace{0.3em}
\noindent\textbf{Analytical component.}
Energy decomposes into three terms.
Cache rebuild energy captures the bulk transfers at window boundaries:
\begin{equation}
\begin{aligned}
E_{\text{rebuild}}(W)
= \sum_{k=1}^{K} \sum_{m \in \mathcal{O}} \Bigl(
& \epsilon_{\text{init}}(m)\,\mathbb{I}[\Delta_{k,m} \neq \varnothing] \\
& {}+ \epsilon_{\text{payload}}\!\left(m, |\Delta_{k,m}|\right)
\Bigr).
\end{aligned}
\label{eq:rebuild}
\end{equation}

Residual miss energy captures on-demand fetches within each window:
\begin{equation}
\begin{aligned}
E_{\text{miss}}(W)
= \sum_{b=1}^{B} \sum_{m \in \mathcal{O}} \Bigl(
& \epsilon_{\text{init}}(m)\,\mathbb{I}[M_{b,m} \neq \varnothing] \\
& {}+ \epsilon_{\text{payload}}\!\left(m, |M_{b,m}|\right)
\Bigr).
\end{aligned}
\label{eq:miss}
\end{equation}

GPU stall energy is the product of the accelerator's idle power draw $P_{\text{idle}}$ and total stall time $T_{\text{stall}}(W)$, where stall time sums across all rebuild and miss events, with each event's duration modeled as the maximum of the slowest single-owner transfer and the aggregate bandwidth ceiling:
\begin{equation}
E_{\text{stall}}(W) = P_{\text{idle}} \cdot T_{\text{stall}}(W).
\label{eq:stall}
\end{equation}

The full analytical estimate is
\[
E_{\text{phy}}(W)
= E_{\text{rebuild}}(W) + E_{\text{miss}}(W) + E_{\text{stall}}(W).
\]
This model exposes the central trade-off: larger windows amortize rebuild initiations but risk stale hot sets with rising residual misses; smaller windows do the opposite.

\vspace{0.3em}
\noindent\textbf{Learned rank correction.}
Real systems exhibit effects the analytical model cannot capture precisely: partial overlap between communication and computation, contention on shared resources, and non-linear bandwidth saturation.
\name accounts for these with a lightweight learned correction $L(W)$.
For a small set of calibration window sizes $\mathcal{W}_{\text{cal}} \subseteq \mathcal{W}$, we run training and measure ground-truth energy $E_{\text{GT}}(W)$.
A linear model predicts the residual: $L(W) = \beta^\top z(W) + \gamma_W$,
where $z(W)$ is a feature vector produced by the simulator (rebuild count, miss count, stall time, hit ratio), $\beta$ is a global weight vector, and $\gamma_W$ is a per-window bias.
The model is trained with a pairwise ranking loss that preserves the correct ordering of window sizes rather than minimizing absolute prediction error (Section~\ref{subsec:impl_calibration}).
The final score used by the optimizer is:
\begin{equation}
E_{\text{net}}(W) = E_{\text{phy}}(W) + L(W).
\label{eq:enet}
\end{equation}
The optimizer selects $W^* = \arg\min_{W \in \mathcal{W}} E_{\text{net}}(W)$.
Because $|\mathcal{W}|$ is small and the simulator operates on a single-epoch trace, the entire offline phase completes in under five seconds.

\subsection{Runtime Execution}
\label{subsec:runtime}

At runtime, \name pipelines cache construction with GNN training so that the GPU is never blocked waiting for a full cache rebuild.
Algorithm~\ref{alg:runtime} gives the complete procedure.
\begin{algorithm}[t]
\caption{\name Pipelined Runtime Execution}
\label{alg:runtime}
\begin{algorithmic}[1]
\Require Presampled trace $\mathcal{T}$, window size $W^*$, hot-set budget $n_{\text{hot}}$
\Statex
\State $C \gets \varnothing$ \Comment{Current cache}
\State Build hot set $H_1$ from $\mathcal{T}$ for window 1
\State \textsc{Prefetcher.WarmCache}($H_1$) \Comment{Begin staging}
\Statex
\For{$k = 1$ \textbf{to} $\lceil B / W^* \rceil$}
  \State $C \gets$ \textsc{Prefetcher.AwaitCache}() \Comment{Block until $H_k$ is ready}
  \Statex
  \If{$k < \lceil B / W^* \rceil$} \Comment{Look-ahead: warm next window}
    \State Build hot set $H_{k+1}$ from $\mathcal{T}$
    \State \textsc{Prefetcher.WarmCache}($H_{k+1} \setminus C$) \Comment{Parallel with training}
  \EndIf
  \Statex
  \For{each batch $b$ in window $k$}
    \State Look up cached features for $N_b \cap H_k$
    \State $M_b \gets N_b \setminus H_k$ \Comment{Residual misses}
    \State \textsc{Prefetcher.FetchMisses}($M_b$) \Comment{Async on-demand}
    \State Await miss results; assemble full feature tensor
    \State Forward pass, backward pass, parameter update
  \EndFor
\EndFor
\end{algorithmic}
\end{algorithm}

There are three aspects of this design. First, the prefetcher
is a dedicated background thread that builds the cache for window
$k{+}1$ while the GPU trains on window $k$'s batches (lines 7--9).
It issues one bulk RPC per partition owner, and because each window
takes tens of seconds to train, the transfers complete well before
the current window finishes. The training loop blocks only briefly at
each window boundary to confirm the cache is ready (line~5).
 
Second, nodes outside the hot set are fetched on demand by the same
prefetcher (lines 12--14), using the same asynchronous mechanism that
baseline DistDGL uses for all remote accesses. \name simply reduces
the volume reaching this path by an order of magnitude.
 
Third, the core energy benefit comes from consolidation. For
$W{=}16$ batches across three remote owners, \name replaces up to
$16{\times}3 = 48$ individual RPCs with 3 bulk transfers, each
carrying hundreds or thousands of features. The initiation cost
$\epsilon_{\text{init}}$ is paid 3 times instead of 48, placing the
system in the payload-dominated regime identified in
Section~\ref{subsec:init_overhead}.
\section{Implementation}
\label{sec:implementation}

We implement \name on top of PyTorch and DistDGL.
This section describes the three components that support the offline phase: the energy profiling infrastructure (Section~\ref{subsec:impl_profiler}), the trace simulator (Section~\ref{subsec:impl_simulator}), and the rank calibration module (Section~\ref{subsec:impl_calibration}).

\subsection{Energy Profiling Infrastructure}
\label{subsec:impl_profiler}

\begin{figure}[t]
  \centering
  \vspace{-15pt}
  \includegraphics[width=\columnwidth]{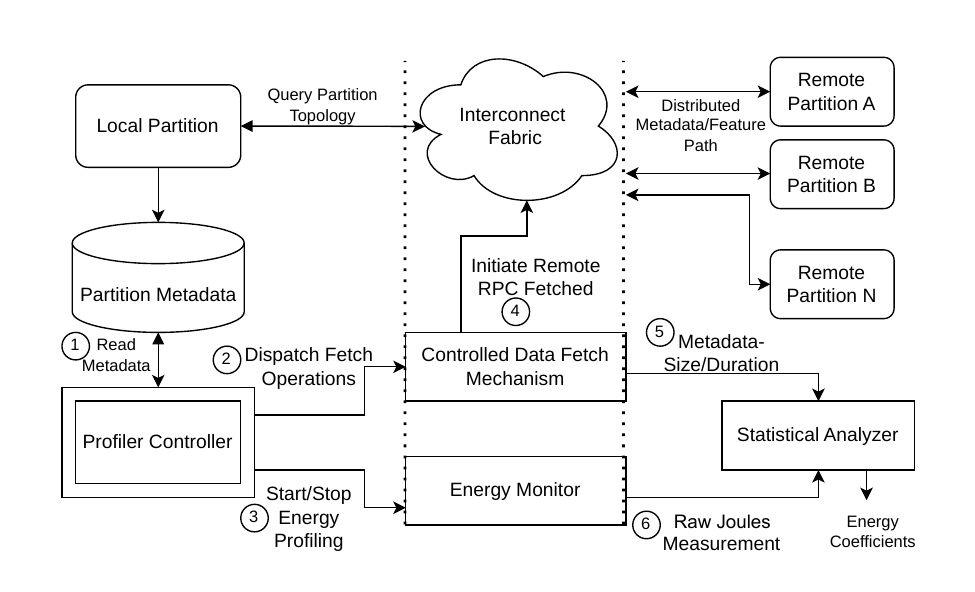}
  \caption{\name energy profiling pipeline. The Profiler Controller reads partition metadata~\cmark{1}, dispatches controlled fetches~\cmark{2}--\cmark{3}, collects transfer metadata~\cmark{4}, and triggers energy measurement~\cmark{5}--\cmark{6}. The Statistical Analyzer derives per-owner energy and latency coefficients from the raw readings.}
  \label{fig:profiling_pipeline}
\end{figure}

The profiling subsystem operates as a lightweight daemon that interfaces with hardware counters to capture energy consumption at millisecond granularity. Figure~\ref{fig:profiling_pipeline} illustrates the six-step profiling pipeline.
 
CPU energy is read through the Intel RAPL interface and GPU power is
queried via NVML, integrated over time to obtain joules. At
initialization, the profiler measures a quiescent baseline
$P_{\text{idle}}$ to separate static power from active operation
costs. Each remote partition owner $m$ is then probed with two types
of synthetic workloads: \emph{initiation probes}, which issue loops
of single-node fetches to isolate the fixed RPC overhead
$\epsilon_{\text{init}}(m)$, and \emph{payload probes}, which sweep
across transfer sizes to characterize
$\epsilon_{\text{payload}}(m, x)$. The known initiation cost is
subtracted from payload measurements so that the residual isolates
per-feature transfer energy. Latency surrogates
$\lambda_{\text{init}}(m)$ and $\lambda_{\text{payload}}(m, x)$ are
derived analogously. A linear fit of
$\epsilon_{\text{payload}}(m, x)$ as a function of transfer size
balances model simplicity with evaluation speed during the window
sweep.

\subsection{Trace Simulator}
\label{subsec:impl_simulator}

The simulator enables offline exploration of window sizes without running actual training.
It consumes the deterministic access trace from the Presampled~DB and the partition book generated by DistDGL.

For a candidate $W$, the simulator makes a single pass over the trace and maintains the cache state across windows.
Two types of events are accumulated:

\noindent\textbf{Rebuild events} occur at window boundaries.
The simulator computes the set difference between the new hot set and the previous cache, groups the delta by owner, and accumulates initiation and payload energy using the profiled coefficients (Equation~\ref{eq:rebuild}).

\noindent\textbf{Miss events} occur within each window.
For every batch, the simulator identifies required nodes absent from the hot set and accumulates per-owner on-demand fetch costs (Equation~\ref{eq:miss}).

In parallel, the simulator tracks stall time using the latency surrogates and multiplies the total by $P_{\text{idle}}$ to obtain GPU stall energy (Equation~\ref{eq:stall}).
The pass produces both the analytical estimate $E_{\text{phy}}(W)$ and the feature vector $z(W)$ consumed by the rank calibration module.

\subsection{Rank Calibration}
\label{subsec:impl_calibration}

The analytical model captures the dominant cost structure but cannot account for all physical effects.
The calibration module corrects for these gaps using a learning-to-rank approach: the goal is to preserve the correct ordering of window sizes, not to predict absolute energy.

\noindent\textbf{Feature extraction.}
For each $W$, the simulator outputs the feature vector $z(W)$ alongside $E_{\text{phy}}(W)$.
A small set of calibration window sizes $\mathcal{W}_{\text{cal}} \subseteq \mathcal{W}$ are run on the actual cluster to obtain ground-truth energy $E_{\text{GT}}(W)$.

\noindent\textbf{Ranking objective.}
A linear model predicts a corrected score $S(W) = \beta^\top z(W) + \gamma_W$, where $\beta$ is a global weight vector and $\gamma_W$ is a per-window bias.
For any pair $(W_i, W_j)$, let $\sigma_{ij} = \mathrm{sign}(E_{\text{GT}}(W_j) - E_{\text{GT}}(W_i))$ encode the true ordering.
We minimize a pairwise ranking loss:
\begin{equation}
\mathcal{L} = \sum_{i < j} \log\!\bigl(1 + \exp\!\bigl(-\sigma_{ij}(S(W_j) - S(W_i))\bigr)\bigr) + \lambda \|\beta\|_2^2,
\label{eq:ranking_loss}
\end{equation}
which penalizes every pair whose predicted order disagrees with the measured order.
The regularization term prevents overfitting to the small calibration set.
Optimization is lightweight (a linear model over a handful of features and seven candidates) and converges in milliseconds.
The resulting weights are reused across training runs on the same hardware.
\section{Evaluation}\label{sec:evaluation}\label{sec:results}

We evaluate \name against four distributed GNN training systems
across three datasets and three batch sizes, measuring total energy,
CPU/GPU energy breakdown, throughput, and simulator fidelity.
\begin{figure}[t]
  \centering
  \includegraphics[width=\columnwidth]{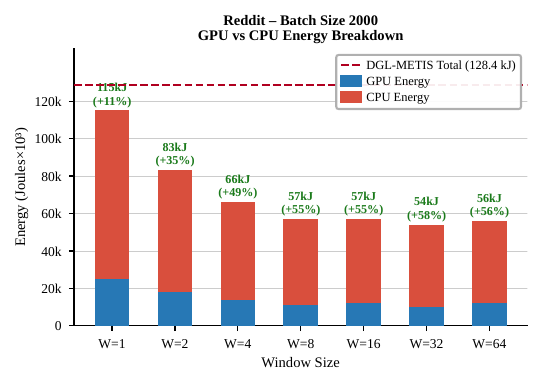}
  \caption{GPU vs.\ CPU energy breakdown for Reddit at $B{=}2000$
  across window sizes. The dashed line marks the DGL-METIS baseline
  (128.4\,kJ). Both components decrease as $W$ increases, with CPU
  energy accounting for most of the reduction.}
  \label{fig:gpu_cpu_breakdown}
\end{figure}

\subsection{Experimental Setup}
\label{sec:setup}

\textbf{Platform and model.}
Experiments run on a 4-node Chameleon Cloud cluster. Each node has an
Intel Xeon CPU, two NVIDIA P100 GPUs, and 25\,Gbps Ethernet. Each
graph is partitioned into 4 METIS partitions, one per node. All
systems train a 2-layer GraphSAGE model with 16 hidden units, fan-out
$\{10,25\}$, learning rate 0.003, dropout 0.5, and 30 epochs. Unless
otherwise noted, we use batch size $B{=}2000$ and also evaluate
$B{=}1000$ and $B{=}3000$.

\textbf{Measurement and baselines.}
CPU energy is measured with Intel RAPL and GPU energy with NVIDIA
NVML, both sampled every 50\,ms. For Default DGL, which does not
support in-process NVML logging, we use an external daemon on each
node. We report the sum across all 4 nodes over the full 30-epoch
run. We compare against \emph{Default DGL}, \emph{BGL}~\cite{liu2023bgl},
\emph{RapidGNN}~\cite{rapidgnn2025}, and
\emph{GraphStorm}~\cite{zheng2024graphstorm}. Relative to RapidGNN,
\name adds window-based bulk transfer consolidation, GPU frequency
reduction during CPU-bound sampling, and a simulator-guided autotuner.

\textbf{Datasets.}
We use Reddit (233K nodes, 114M edges), OGBN-Products (2.4M nodes,
61.9M edges), and OGBN-Papers100M (111M nodes, 1.6B edges).

\subsection{Window Size Analysis}
\label{subsec:window_analysis}

We first validate the central design assumption: window-based caching
produces a convex energy curve over $W$ and reduces both CPU and GPU
energy.

Figure~\ref{fig:gpu_cpu_breakdown} shows the CPU/GPU energy breakdown
for Reddit at $B{=}2000$. The DGL-METIS baseline (128.4\,kJ, dashed
line) uses standard on-demand feature fetching. CPU energy accounts
for 78--85\% of total energy and drops from about 90\,kJ at $W{=}1$
to 46\,kJ at $W{=}32$ as bulk transfers replace thousands of small
RPCs. GPU energy follows the same trend, falling from about 25\,kJ to
8\,kJ as larger windows reduce blocking communication and GPU stall
time. Frequency scaling during sampling further lowers idle GPU power.

Total energy decreases steadily from $W{=}1$ to $W{=}32$, then
flattens at $W{=}64$ as hot-set staleness begins to offset further
amortization. The minimum is $W{=}32$ at 54\,kJ, a 58\% reduction
from baseline. Windows in $\{8,16,32,64\}$ remain within 3
percentage points of this optimum, confirming that the energy valley
is broad enough to make near-optimal tuning practical.

\begin{figure}[t]
  \centering
  \includegraphics[width=\columnwidth]{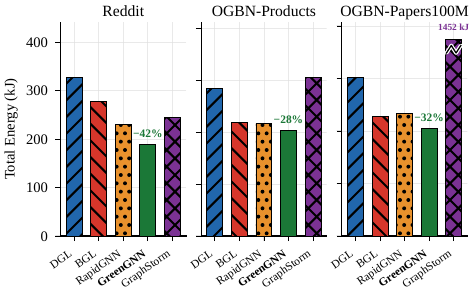}
  \caption{Total energy consumption (GPU + CPU, all nodes) across
  three datasets at $B{=}2000$. \name achieves the lowest energy in
  all cases. Annotations show percentage reduction relative to Default
  DGL. GraphStorm on Papers100M is truncated (actual: 1,452\,kJ).}
  \label{fig:total_energy}
\end{figure}
\subsection{End-to-End Energy Consumption}\label{sec:total_energy}

Figure~\ref{fig:total_energy} compares total system energy (GPU +
CPU, all nodes) at $B{=}2000$. \name achieves the lowest total
energy on every dataset: 189.4\,kJ on Reddit (42.0\% below Default
DGL's 326.8\,kJ), 203.9\,kJ on OGBN-Products (28.3\% below
284.3\,kJ), and 307.2\,kJ on OGBN-Papers100M (32.1\% below
452.2\,kJ). In absolute terms, these correspond to savings of
137.4\,kJ, 80.4\,kJ, and 145.0\,kJ per training run. At cloud
scale, where training jobs run continuously across hundreds of GPUs,
per-run savings of this magnitude compound into meaningful reductions
in operational cost and carbon footprint.

The savings stem from two mechanisms formalized in
Section~\ref{subsec:cost}. Window-based caching converts thousands
of fine-grained per-batch RPCs into a small number of bulk transfers
per window, reducing accumulated initiation energy $E_{\text{rebuild}}$
(Equation~\ref{eq:rebuild}). Bulk transfers also shorten the
intervals during which GPUs stall on remote data, reducing
$P_{\text{idle}} \cdot T_{\text{stall}}$
(Equation~\ref{eq:stall}). The 6.2--17.8\% improvement
over RapidGNN reflects the added contribution of GPU frequency
reduction during sampling, bulk transfer consolidation, and the
energy-aware window selection by the autotuner.

GraphStorm presents a notable contrast. On smaller datasets it
remains competitive (245.3\,kJ on Reddit, 305.0\,kJ on Products),
but on OGBN-Papers100M it consumes 1,452\,kJ, $4.7\times$ the
energy of \name. This increase is driven by post-training overhead
(model serialization, distributed inference setup) whose cost grows
super-linearly with graph size. The truncated bar in
Figure~\ref{fig:total_energy} annotates GraphStorm's actual value.

\subsection{GPU Energy Analysis}\label{sec:gpu_energy}

CPU-side costs (data loading, sampling, RPC handling) account for
77--95\% of total energy across all frameworks. Isolating the GPU
component reveals the effect of \name's two GPU-targeted mechanisms:
reduced stall time from bulk transfer consolidation and reduced idle
power from frequency scaling during sampling.

\begin{figure}[t]
  \centering
  \includegraphics[width=\columnwidth]{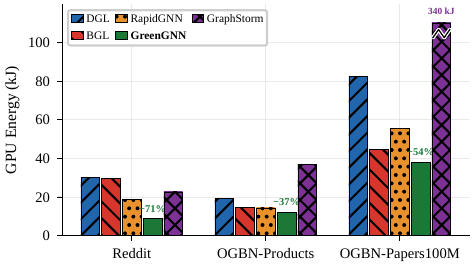}
  \caption{GPU energy at $B{=}2000$. \name achieves 37--71\% GPU
  energy reduction over Default DGL. GraphStorm on Papers100M is
  truncated (actual: 340\,kJ).}
  \label{fig:gpu_energy}
\end{figure}

Figure~\ref{fig:gpu_energy} shows GPU-only energy at $B{=}2000$.
\name reduces GPU energy by 70.9\% on Reddit (8.8\,kJ vs.\
30.2\,kJ for DGL), 37.5\% on OGBN-Products (12.0\,kJ vs.\
19.2\,kJ), and 54.0\% on OGBN-Papers100M (38.0\,kJ vs.\
82.6\,kJ). These GPU reductions are disproportionately larger than
the total-energy reductions (37--71\% vs.\ 28--42\%) because
consolidating remote accesses into bulk window transfers directly
reduces $T_{\text{stall}}$, and GPU power draw remains high even
when idle. The frequency reduction during CPU-bound sampling phases
compounds this effect by lowering $P_{\text{idle}}$ itself when the
GPU is not needed.

Even compared to RapidGNN, which shares the same presampling
infrastructure and achieves similar reductions in remote fetch volume,
\name delivers an additional 16.1--52.7\% GPU energy reduction. This
gap isolates the contribution of the energy-aware components unique
to \name: GPU frequency management during sampling and
simulator-guided window selection that picks $W^*$ to minimize energy
rather than maximize throughput. GraphStorm consumes 340.4\,kJ of
GPU energy on Papers100M ($4.1\times$ that of Default DGL),
reflecting sustained high GPU power draw during its post-training
phases.

\subsection{Energy and Throughput}\label{sec:pareto}

\begin{figure}[t]
  \centering
  \includegraphics[width=\columnwidth]{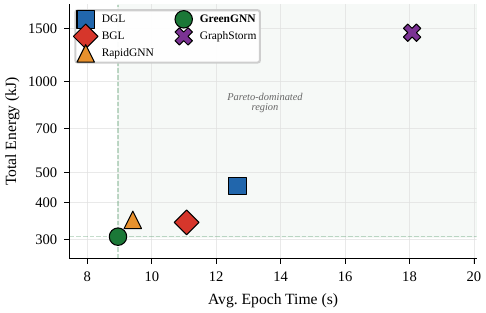}
  \caption{Energy vs.\ epoch time on OGBN-Papers100M at $B{=}2000$
  (log-scale $y$-axis). \name is Pareto-optimal. The shaded region
  denotes the Pareto-dominated area.}
  \label{fig:pareto}
\end{figure}

A natural question is whether energy savings come at the expense of
training speed. Figure~\ref{fig:pareto} plots each framework in
energy vs.\ epoch-time space on OGBN-Papers100M at $B{=}2000$.
\name occupies the Pareto-optimal position, simultaneously achieving
the lowest total energy (307.2\,kJ) and the fastest average epoch
time (8.95\,s). All other frameworks are Pareto-dominated: RapidGNN
is 13.7\% more energy-intensive and 5.1\% slower, BGL 11.6\% more
energy-intensive and 23.8\% slower, Default DGL 47.2\% more
energy-intensive and 41.5\% slower, and GraphStorm $4.7\times$ the
energy and $2.0\times$ slower.

This result may seem counterintuitive, since one might expect GPU
frequency reduction to degrade throughput. \name avoids this tradeoff
because frequency is reduced only during the CPU-bound sampling
phase, when the GPU is idle regardless. During forward and backward
passes the GPU runs at full frequency. The window-based caching
mechanism further improves throughput by reducing blocking RPC
round-trips per batch. The net effect is that bulk transfer
consolidation improves both energy and training time simultaneously,
because the same mechanism that reduces initiation energy also
reduces the communication bottleneck that limits throughput.

\subsection{Epoch-Time Speedup}\label{sec:speedup}

\begin{figure}[t]
  \centering
  \includegraphics[width=\columnwidth]{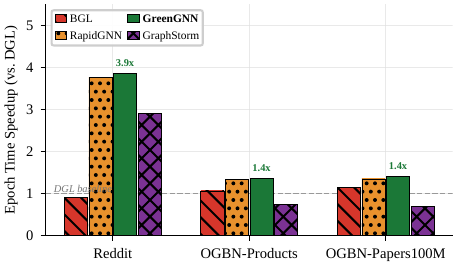}
  \caption{Epoch-time speedup relative to Default DGL at $B{=}2000$.
  The dashed line marks the DGL baseline ($1.0\times$).}
  \label{fig:speedup}
\end{figure}

Figure~\ref{fig:speedup} quantifies per-epoch speedup relative to
Default DGL at $B{=}2000$. \name achieves $3.9\times$ on Reddit
(1.77\,s vs.\ 6.84\,s), $1.4\times$ on Products (2.77\,s vs.\
3.80\,s), and $1.4\times$ on Papers100M (8.95\,s vs.\ 12.66\,s).
RapidGNN achieves similar speedups ($3.8\times$, $1.3\times$,
$1.3\times$), confirming that \name's energy-aware components
introduce no throughput overhead on top of the shared presampling
mechanism. The larger gain on Reddit reflects its smaller graph and
higher locality, which allow the window cache to absorb a larger
fraction of remote requests.

BGL offers negligible speedup ($0.9$--$1.1\times$), consistent with
its focus on I/O optimization rather than communication reduction.
GraphStorm achieves $2.9\times$ on Reddit, where its distributed
sampling amortizes well over the small graph, but falls below the DGL
baseline on Products ($0.7\times$) and Papers100M ($0.7\times$). On
larger graphs, its coordination overhead for distributed sampling,
model synchronization, and post-training serialization exceeds the
parallelism benefit.

\subsection{Robustness Across Batch Sizes}\label{sec:heatmap}

\begin{figure}[t]
  \centering
  \includegraphics[width=\columnwidth]{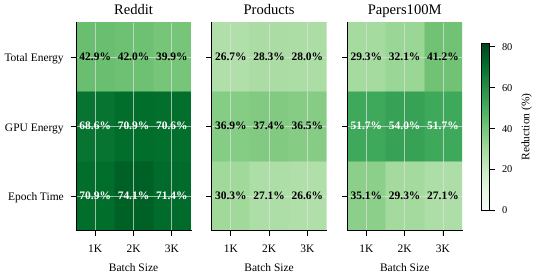}
  \caption{\name's reduction over Default DGL (\%) across three
  metrics and three batch sizes per dataset. All cells are positive.
  GPU energy reductions (36--71\%) are uniformly the largest.}
  \label{fig:heatmap}
\end{figure}

Figure~\ref{fig:heatmap} presents \name's percentage reduction over
Default DGL across three metrics (total energy, GPU energy, epoch
time) and three batch sizes ($B \in \{1000, 2000, 3000\}$) for each
dataset. Every cell is positive: \name improves on all metrics in all
27 configurations.

GPU energy reductions are the largest and most consistent, ranging
from 36.4\% (Products, $B{=}3000$) to 70.9\% (Reddit, $B{=}2000$),
reflecting the effectiveness of bulk transfer consolidation at
reducing GPU stall time. Total energy reductions range from 26.7\% to
42.9\%, with even the smallest (Products at $B{=}1000$) representing
a 78.4\,kJ saving. On Papers100M, total energy reduction increases
with batch size (29.2\% $\to$ 32.1\% $\to$ 41.2\% for
$B{=}$1000/2000/3000), consistent with the cost model: larger
batches produce more remote feature requests per window, so bulk
transfer consolidation amortizes $E_{\text{init}}$ over a larger
payload, pushing the system further into the payload-dominated regime
(Section~\ref{subsec:init_overhead}).

\subsection{Comprehensive Results}\label{sec:table}
\begin{table*}[t]
\centering
\caption{Comprehensive energy and epoch time across all frameworks, datasets,
and batch sizes. \textbf{Bold} = best (lowest) per column.
$\Delta$ = reduction vs.\ Default DGL. All energy values in kJ.}
\label{tab:comprehensive}
\renewcommand{\arraystretch}{1.05}
\scriptsize
\begin{subtable}{\textwidth}
\centering
\caption{Reddit}
\begin{tabular}{l|rrrr|rrrr|rrrr}
\hline
 & \multicolumn{4}{c|}{\textbf{$B{=}1000$}} & \multicolumn{4}{c|}{\textbf{$B{=}2000$}} & \multicolumn{4}{c}{\textbf{$B{=}3000$}} \\
\textbf{Framework} & GPU & CPU & Total & ET(s) & GPU & CPU & Total & ET(s) & GPU & CPU & Total & ET(s) \\
\hline
DGL & 41.2 & 331.2 & 372.4 & 9.20 & 30.2 & 296.6 & 326.8 & 6.84 & 25.8 & 281.9 & 307.7 & 5.67 \\
BGL & 38.2 & 288.4 & 326.6 & 9.68 & 29.4 & 248.6 & 278.0 & 7.47 & 22.8 & 228.3 & 251.1 & 5.77 \\
RapidGNN & 27.1 & 239.4 & 266.5 & \textbf{2.65} & 18.6 & 211.8 & 230.4 & 1.82 & 14.2 & 198.3 & 212.5 & 1.66 \\
GreenGNN & \textbf{12.9} & \textbf{199.7} & \textbf{212.6} & 2.68 & \textbf{8.8} & \textbf{180.6} & \textbf{189.4} & \textbf{1.77} & \textbf{7.6} & \textbf{177.5} & \textbf{185.1} & \textbf{1.62} \\
GraphStorm & 29.4 & 260.3 & 289.6 & 3.71 & 22.6 & 222.7 & 245.3 & 2.36 & 21.9 & 220.1 & 242.0 & 2.21 \\
\hline
$\Delta$ GreenGNN & \textcolor{green!60!black}{-69\%} & \textcolor{green!60!black}{-40\%} & \textcolor{green!60!black}{-43\%} & \textcolor{green!60!black}{-71\%} & \textcolor{green!60!black}{-71\%} & \textcolor{green!60!black}{-39\%} & \textcolor{green!60!black}{-42\%} & \textcolor{green!60!black}{-74\%} & \textcolor{green!60!black}{-71\%} & \textcolor{green!60!black}{-37\%} & \textcolor{green!60!black}{-40\%} & \textcolor{green!60!black}{-71\%} \\
\hline
\end{tabular}
\end{subtable}
\vspace{4pt}

\begin{subtable}{\textwidth}
\centering
\caption{OGBN-Products}
\begin{tabular}{l|rrrr|rrrr|rrrr}
\hline
 & \multicolumn{4}{c|}{\textbf{$B{=}1000$}} & \multicolumn{4}{c|}{\textbf{$B{=}2000$}} & \multicolumn{4}{c}{\textbf{$B{=}3000$}} \\
\textbf{Framework} & GPU & CPU & Total & ET(s) & GPU & CPU & Total & ET(s) & GPU & CPU & Total & ET(s) \\
\hline
DGL & 20.7 & 273.2 & 293.8 & 4.26 & 19.2 & 265.1 & 284.3 & 3.80 & 16.2 & 256.4 & 272.6 & 3.20 \\
BGL & 16.3 & 219.5 & 235.8 & 3.98 & 14.5 & 204.2 & 218.8 & 3.58 & 13.2 & 200.2 & 213.4 & 3.25 \\
RapidGNN & 16.8 & 217.6 & 234.4 & 3.13 & 14.3 & 203.1 & 217.4 & 2.86 & 12.1 & 198.4 & 210.5 & \textbf{2.26} \\
GreenGNN & \textbf{13.1} & \textbf{202.3} & \textbf{215.4} & \textbf{2.97} & \textbf{12.0} & \textbf{191.8} & \textbf{203.9} & \textbf{2.77} & \textbf{10.3} & \textbf{186.1} & \textbf{196.3} & 2.35 \\
GraphStorm & 46.0 & 250.3 & 296.3 & 6.32 & 36.7 & 268.4 & 305.0 & 5.19 & 33.8 & 257.8 & 291.6 & 4.33 \\
\hline
$\Delta$ GreenGNN & \textcolor{green!60!black}{-37\%} & \textcolor{green!60!black}{-26\%} & \textcolor{green!60!black}{-27\%} & \textcolor{green!60!black}{-30\%} & \textcolor{green!60!black}{-37\%} & \textcolor{green!60!black}{-28\%} & \textcolor{green!60!black}{-28\%} & \textcolor{green!60!black}{-27\%} & \textcolor{green!60!black}{-37\%} & \textcolor{green!60!black}{-27\%} & \textcolor{green!60!black}{-28\%} & \textcolor{green!60!black}{-27\%} \\
\hline
\end{tabular}
\end{subtable}
\vspace{4pt}

\begin{subtable}{\textwidth}
\centering
\caption{OGBN-Papers100M}
\begin{tabular}{l|rrrr|rrrr|rrrr}
\hline
 & \multicolumn{4}{c|}{\textbf{$B{=}1000$}} & \multicolumn{4}{c|}{\textbf{$B{=}2000$}} & \multicolumn{4}{c}{\textbf{$B{=}3000$}} \\
\textbf{Framework} & GPU & CPU & Total & ET(s) & GPU & CPU & Total & ET(s) & GPU & CPU & Total & ET(s) \\
\hline
DGL & 89.5 & 395.5 & 484.9 & 15.78 & 82.6 & 369.6 & 452.2 & 12.66 & 71.2 & 422.7 & 494.0 & 11.09 \\
BGL & 53.4 & \textbf{273.0} & \textbf{326.3} & 13.35 & 44.6 & 298.1 & 342.7 & 11.08 & 44.6 & 295.9 & 340.6 & 11.09 \\
RapidGNN & 67.3 & 335.2 & 402.5 & 11.33 & 55.4 & 293.8 & 349.2 & 9.41 & 48.2 & 278.4 & 326.6 & 8.43 \\
GreenGNN & \textbf{43.2} & 299.9 & 343.1 & \textbf{10.24} & \textbf{38.0} & \textbf{269.2} & \textbf{307.2} & \textbf{8.95} & \textbf{34.4} & \textbf{255.9} & \textbf{290.3} & \textbf{8.09} \\
GraphStorm & 440.6 & 1463.5 & 1904.1 & 26.15 & 340.4 & 1111.5 & 1451.9 & 18.08 & 361.2 & 1163.5 & 1524.7 & 16.46 \\
\hline
$\Delta$ GreenGNN & \textcolor{green!60!black}{-52\%} & \textcolor{green!60!black}{-24\%} & \textcolor{green!60!black}{-29\%} & \textcolor{green!60!black}{-35\%} & \textcolor{green!60!black}{-54\%} & \textcolor{green!60!black}{-27\%} & \textcolor{green!60!black}{-32\%} & \textcolor{green!60!black}{-29\%} & \textcolor{green!60!black}{-52\%} & \textcolor{green!60!black}{-39\%} & \textcolor{green!60!black}{-41\%} & \textcolor{green!60!black}{-27\%} \\
\hline
\end{tabular}
\end{subtable}
\vspace{4pt}

\end{table*}

Table~\ref{tab:comprehensive} consolidates GPU energy, CPU energy,
total energy, and average epoch time for all five frameworks across
every dataset and batch size.

Four observations emerge. First, \name achieves the lowest total
energy in 8 of 9 configurations. The exception is Papers100M at
$B{=}1000$, where BGL is 5.1\% lower due to aggressive host-memory
caching on that partition layout. Even there, \name's GPU energy is
19.1\% lower than BGL's, indicating that the window-based mechanism
is more effective at reducing accelerator stall waste while BGL's
advantage comes from CPU-side caching.

Second, \name achieves the lowest or tied-lowest epoch time in 7 of
9 configurations, confirming that energy savings do not degrade
throughput. The two exceptions (Reddit at $B{=}1000$ and Products at
$B{=}3000$) show epoch times within 4\% of the fastest framework.

Third, GraphStorm is consistently the most energy-intensive framework
on Papers100M, consuming 1,452--1,905\,kJ depending on batch size
($4.7$--$5.5\times$ the energy of \name). This illustrates the cost
of frameworks that optimize for feature richness without considering
the energy implications of their communication design.

Fourth, the 6.2--20.2\% gap between \name and RapidGNN in total
energy isolates the contribution of GPU frequency reduction and
simulator-guided window selection. This translates to
13.5--59.4\,kJ per run, savings that accumulate in production
settings where models are retrained daily across multiple
hyperparameter configurations.

\subsection{Simulator Validation}\label{subsec:eval_simulator}
The autotuner's value depends on whether its predicted window
rankings match the true energy rankings on the cluster.
Table~\ref{tab:simulator_ranking} reports the simulator-predicted
rank (S) and measured rank (A) for each window size across all nine
configurations on Reddit, OGBN-Products, and OGBN-Papers100M.

\begin{table}[!htb]
\centering
\caption{Simulator ranking fidelity. For each window size, the
simulator's predicted rank (S) and measured rank (A) are shown.
$\checkmark$ marks correct top-1 predictions. $\tau$ is Kendall's
rank correlation.}
\label{tab:simulator_ranking}
\footnotesize
\setlength{\tabcolsep}{2.5pt}
\begin{tabular}{ll cc cc cc cc cc cc cc  c}
\toprule
& & \multicolumn{2}{c}{$W{=}1$} & \multicolumn{2}{c}{$W{=}2$}
& \multicolumn{2}{c}{$W{=}4$} & \multicolumn{2}{c}{$W{=}8$}
& \multicolumn{2}{c}{$W{=}16$} & \multicolumn{2}{c}{$W{=}32$}
& \multicolumn{2}{c}{$W{=}64$} & \\
\cmidrule(lr){3-4}\cmidrule(lr){5-6}\cmidrule(lr){7-8}
\cmidrule(lr){9-10}\cmidrule(lr){11-12}\cmidrule(lr){13-14}
\cmidrule(lr){15-16}
& $B$
& S & A & S & A & S & A & S & A & S & A & S & A & S & A
& $\tau$ \\
\midrule
\multirow{3}{*}{Red.}
& 1k & 7 & 7 & 6 & 6 & 5 & 5 & 2 & 4 & \textbf{1} & \textbf{1}$^{\checkmark}$ & 3 & 2 & 4 & 3 & 0.81 \\
& 2k & 7 & 7 & 6 & 6 & 5 & 5 & 2 & 4 & \textbf{1} & 3 & 3 & \textbf{1} & 4 & 2 & 0.62 \\
& 3k & 7 & 7 & 6 & 6 & 5 & 5 & 2 & \textbf{1} & \textbf{1} & 4 & 3 & 2 & 4 & 3 & 0.71 \\
\midrule
\multirow{3}{*}{Prod.}
& 1k & 7 & 7 & 6 & 6 & 5 & 5 & 2 & 3 & \textbf{1} & \textbf{1}$^{\checkmark}$ & 3 & 2 & 4 & 4 & 0.90 \\
& 2k & 7 & 7 & 6 & 6 & 5 & 5 & 2 & 2 & \textbf{1} & \textbf{1}$^{\checkmark}$ & 3 & 3 & 4 & 4 & 1.00 \\
& 3k & 7 & 7 & 6 & 6 & 5 & 5 & 2 & 3 & \textbf{1} & \textbf{1}$^{\checkmark}$ & 3 & 2 & 4 & 4 & 0.90 \\
\midrule
\multirow{3}{*}{Pap.}
& 1k & 7 & 7 & 6 & 6 & 5 & 5 & 2 & 3 & \textbf{1} & 2 & 3 & \textbf{1} & 4 & 4 & 0.81 \\
& 2k & 7 & 7 & 6 & 6 & 5 & 5 & 2 & 3 & \textbf{1} & \textbf{1}$^{\checkmark}$ & 3 & 2 & 4 & 4 & 0.90 \\
& 3k & 7 & 7 & 6 & 6 & 5 & 5 & 2 & 2 & \textbf{1} & \textbf{1}$^{\checkmark}$ & 3 & 3 & 4 & 4 & 1.00 \\
\bottomrule
\end{tabular}
\end{table}

The simulator achieves 6/9 top-1 accuracy, correctly identifying
$W{=}16$ as optimal for Reddit $B{=}1000$, for OGBN-Products at all
three batch sizes, and for OGBN-Papers100M at $B{=}2000$ and
$B{=}3000$. In the three configurations where the top pick differs
from the true optimum (Reddit $B{=}2000$ and $B{=}3000$;
Papers100M $B{=}1000$), the true optimum still falls within the
simulator's top-3 candidates, and the energy penalty of selecting the
simulator's pick is at most 5.3\% (3\,kJ on Reddit) and 4.8\%
(8\,kJ on Papers100M).

Kendall's $\tau$ ranges from 0.62 to 1.00 across the nine
configurations (mean 0.85), with perfect ordering ($\tau{=}1.00$) on
OGBN-Products $B{=}2000$ and OGBN-Papers100M $B{=}3000$. The
lowest $\tau$ (0.62, Reddit $B{=}2000$) results from swapping
$W{=}16$ and $W{=}32$, whose measured energies differ by only
5.3\%. On Papers100M, the larger graph produces more remote
accesses per window, which strengthens the simulator's ability to
distinguish between window sizes: the mean $\tau$ across Papers100M
configurations is 0.90, the highest of the three datasets. The large differences are driven
by RPC initiation counts, while the smaller differences within the
plateau depend on system-level effects that the learned ranker only
partially corrects.

Because GreenGNN does not modify the model architecture, sampling logic, or gradient computation, model accuracy is identical to baseline DistDGL, we refer readers to~\cite{rapidgnn2025} for detailed accuracy validation.

\section{Conclusion and Future Work}

This paper presented \name, an energy-aware distributed GNN training system that consolidates thousands of fine-grained per-batch RPCs into a small number of per-window bulk transfers. The key insight is that GNN neighbor sampling exhibits bursty temporal locality, where a compact hot set dominates remote accesses over a short window of consecutive mini-batches and then quickly falls out of relevance. \name refreshes the cache at window boundaries rather than tracking every access, and an offline autotuner selects the best window size by replaying a deterministic access trace through a discrete-event simulator. Across benchmark datasets on a 4-node GPU cluster, \name reduces total system energy by 27--43\% and GPU energy by 36--71\% over on-demand DistDGL while achieving 1.4--3.9$\times$ epoch-time speedup. The energy-window curve is consistently convex, and the autotuner achieves Kendall's $\tau$ up to 1.00.

Several directions remain open. The current energy model assumes homogeneous hardware, and extending it to clusters with mixed GPU generations would broaden applicability. The window size is selected once before training, so an adaptive scheme that adjusts $W$ across epochs as access patterns shift could capture additional savings. Integrating RDMA-aware bulk transfer paths could further reduce initiation costs on high-performance interconnects. Finally, evaluating \name on full-graph training methods and GNN architectures beyond GraphSAGE would test the generality of window-based caching more broadly.

\bibliographystyle{IEEEtran}
\bibliography{references}

@inproceedings{fan2019graph,
  title={Graph neural networks for social recommendation},
  author={Fan, Wenqi and Ma, Yao and Li, Qing and He, Yuan and Zhao, Eric and Tang, Jiliang and Yin, Dawei},
  booktitle={The world wide web conference},
  pages={417--426},
  year={2019}
}

@inproceedings{dou2020enhancing,
  title={Enhancing graph neural network-based fraud detectors against camouflaged fraudsters},
  author={Dou, Yingtong and Liu, Zhiwei and Sun, Li and Deng, Yutong and Peng, Hao and Yu, Philip S},
  booktitle={Proceedings of the 29th ACM international conference on information \& knowledge management},
  pages={315--324},
  year={2020}
}

@article{zhang2021bio,
  title={Graph neural networks and their current applications in bioinformatics},
  author={Zhang, Xiao-Meng and Liang, Li and Liu, Lin and Tang, Ming-Jing},
  journal={Frontiers in genetics},
  volume={12},
  pages={690049},
  year={2021},
  publisher={Frontiers Media SA}
}

@article{wu2020comprehensive,
  title={A comprehensive survey on graph neural networks},
  author={Wu, Zonghan and Pan, Shirui and Chen, Fengwen and Long, Guodong and Zhang, Chengqi and Yu, Philip S},
  journal={IEEE transactions on neural networks and learning systems},
  volume={32},
  number={1},
  pages={4--24},
  year={2020},
  publisher={IEEE}
}

@article{hamilton2017inductive,
  title={Inductive representation learning on large graphs},
  author={Hamilton, Will and Ying, Zhitao and Leskovec, Jure},
  journal={Advances in neural information processing systems},
  volume={30},
  year={2017}
}

@article{leskovec2016snap,
  title={Snap: A general-purpose network analysis and graph-mining library},
  author={Leskovec, Jure and Sosi{\v{c}}, Rok},
  journal={ACM Transactions on Intelligent Systems and Technology (TIST)},
  volume={8},
  number={1},
  pages={1--20},
  year={2016},
  publisher={ACM New York, NY, USA}
}

@article{ching2015one,
  title={One trillion edges: Graph processing at facebook-scale},
  author={Ching, Avery and Edunov, Sergey and Kabiljo, Maja and Logothetis, Dionysios and Muthukrishnan, Sambavi},
  journal={Proceedings of the VLDB Endowment},
  volume={8},
  number={12},
  pages={1804--1815},
  year={2015},
  publisher={VLDB Endowment}
}

@article{besta2024parallel,
  title={Parallel and distributed graph neural networks: An in-depth concurrency analysis},
  author={Besta, Maciej and Hoefler, Torsten},
  journal={IEEE Transactions on Pattern Analysis and Machine Intelligence},
  volume={46},
  number={5},
  pages={2584--2606},
  year={2024},
  publisher={IEEE}
}

@inproceedings{zheng2020distdgl,
  title={DistDGL: Distributed graph neural network training for billion-scale graphs},
  author={Zheng, Da and Ma, Chao and Wang, Minjie and Zhou, Jinjing and Su, Qidong and Song, Xiang and Gan, Quan and Zhang, Zheng and Karypis, George},
  booktitle={2020 IEEE/ACM 10th Workshop on Irregular Applications: Architectures and Algorithms (IA3)},
  pages={36--44},
  year={2020},
  organization={IEEE}
}

@inproceedings{gandhi2021p3,
  title={P3: Distributed deep graph learning at scale},
  author={Gandhi, Swapnil and Iyer, Anand Padmanabha},
  booktitle={15th $\{$USENIX$\}$ Symposium on Operating Systems Design and Implementation ($\{$OSDI$\}$ 21)},
  pages={551--568},
  year={2021}
}

@inproceedings{wan2022dgs,
  title={Dgs: Communication-efficient graph sampling for distributed gnn training},
  author={Wan, Xinchen and Chen, Kai and Zhang, Yiming},
  booktitle={2022 IEEE 30th International Conference on Network Protocols (ICNP)},
  pages={1--11},
  year={2022},
  organization={IEEE}
}

@article{shao2024distributed,
  title={Distributed graph neural network training: A survey},
  author={Shao, Yingxia and Li, Hongzheng and Gu, Xizhi and Yin, Hongbo and Li, Yawen and Miao, Xupeng and Zhang, Wentao and Cui, Bin and Chen, Lei},
  journal={ACM Computing Surveys},
  volume={56},
  number={8},
  pages={1--39},
  year={2024},
  publisher={ACM New York, NY}
}

@inproceedings{liu2023bgl,
  title={$\{$BGL$\}$:$\{$GPU-Efficient$\}$$\{$GNN$\}$ training by optimizing graph data $\{$I/O$\}$ and preprocessing},
  author={Liu, Tianfeng and Chen, Yangrui and Li, Dan and Wu, Chuan and Zhu, Yibo and He, Jun and Peng, Yanghua and Chen, Hongzheng and Chen, Hongzhi and Guo, Chuanxiong},
  booktitle={20th USENIX Symposium on Networked Systems Design and Implementation (NSDI 23)},
  pages={103--118},
  year={2023}
}

@inproceedings{sun2023legion,
  title={Legion: Automatically pushing the envelope of $\{$Multi-GPU$\}$ system for $\{$Billion-Scale$\}$$\{$GNN$\}$ training},
  author={Sun, Jie and Su, Li and Shi, Zuocheng and Shen, Wenting and Wang, Zeke and Wang, Lei and Zhang, Jie and Li, Yong and Yu, Wenyuan and Zhou, Jingren and others},
  booktitle={2023 USENIX Annual Technical Conference (USENIX ATC 23)},
  pages={165--179},
  year={2023}
}

@article{karypis1998fast,
  title={A fast and high quality multilevel scheme for partitioning irregular graphs},
  author={Karypis, George and Kumar, Vipin},
  journal={SIAM Journal on scientific Computing},
  volume={20},
  number={1},
  pages={359--392},
  year={1998},
  publisher={SIAM}
}

@inproceedings{lin2020pagraph,
  title={Pagraph: Scaling gnn training on large graphs via computation-aware caching},
  author={Lin, Zhiqi and Li, Cheng and Miao, Youshan and Liu, Yunxin and Xu, Yinlong},
  booktitle={Proceedings of the 11th ACM Symposium on Cloud Computing},
  pages={401--415},
  year={2020}
}

@inproceedings{yang2022gnnlab,
  title={GNNLab: a factored system for sample-based GNN training over GPUs},
  author={Yang, Jianbang and Tang, Dahai and Song, Xiaoniu and Wang, Lei and Yin, Qiang and Chen, Rong and Yu, Wenyuan and Zhou, Jingren},
  booktitle={Proceedings of the Seventeenth European Conference on Computer Systems},
  pages={417--434},
  year={2022}
}

@inproceedings{backstrom2012four,
  title={Four degrees of separation},
  author={Backstrom, Lars and Boldi, Paolo and Rosa, Marco and Ugander, Johan and Vigna, Sebastiano},
  booktitle={Proceedings of the 4th annual ACM Web science conference},
  pages={33--42},
  year={2012}
}

@inproceedings{zhang2023two,
  title={Two-level graph caching for expediting distributed GNN training},
  author={Zhang, Zhe and Luo, Ziyue and Wu, Chuan},
  booktitle={IEEE INFOCOM 2023-IEEE Conference on Computer Communications},
  pages={1--10},
  year={2023},
  organization={IEEE}
}

@misc{nvidia2022h100,
  title={H100 tensor core GPU architecture overview},
  author={NVIDIA, NVIDIA},
  year={2022}
}

@inproceedings{cai2021dgcl,
  title={DGCL: An efficient communication library for distributed GNN training},
  author={Cai, Zhenkun and Yan, Xiao and Wu, Yidi and Ma, Kaihao and Cheng, James and Yu, Fan},
  booktitle={Proceedings of the Sixteenth European Conference on Computer Systems},
  pages={130--144},
  year={2021}
}

@inproceedings{song2024granndis,
  title={GraNNDis: Fast distributed graph neural network training framework for multi-server clusters},
  author={Song, Jaeyong and Jang, Hongsun and Lim, Hunseong and Jung, Jaewon and Kim, Youngsok and Lee, Jinho},
  booktitle={Proceedings of the 2024 International Conference on Parallel Architectures and Compilation Techniques},
  pages={91--107},
  year={2024}
}

@inproceedings{ying2018graph,
  title={Graph convolutional neural networks for web-scale recommender systems},
  author={Ying, Rex and He, Ruining and Chen, Kaifeng and Eksombatchai, Pong and Hamilton, William L and Leskovec, Jure},
  booktitle={Proceedings of the 24th ACM SIGKDD international conference on knowledge discovery \& data mining},
  pages={974--983},
  year={2018}
}

@article{wan2022pipegcn,
  title={Pipegcn: Efficient full-graph training of graph convolutional networks with pipelined feature communication},
  author={Wan, Cheng and Li, Youjie and Wolfe, Cameron R and Kyrillidis, Anastasios and Kim, Nam Sung and Lin, Yingyan},
  journal={arXiv preprint arXiv:2203.10428},
  year={2022}
}

@article{tang2024xgnn,
  title={Xgnn: Boosting multi-gpu gnn training via global gnn memory store},
  author={Tang, Dahai and Wang, Jiali and Chen, Rong and Wang, Lei and Yu, Wenyuan and Zhou, Jingren and Li, Kenli},
  journal={Proceedings of the VLDB Endowment},
  volume={17},
  number={5},
  pages={1105--1118},
  year={2024},
  publisher={VLDB Endowment}
}

@article{rapidgnn2025,
  title={RapidGNN: Communication Efficient Large-Scale Distributed Training of Graph Neural Networks},
  author={Niam, Arefin and Nine, MSQ},
  journal={arXiv preprint arXiv:2505.10806},
  year={2025}
}

@inproceedings{zheng2024graphstorm,
  title={GraphStorm: all-in-one graph machine learning framework for industry applications},
  author={Zheng, Da and Song, Xiang and Zhu, Qi and Zhang, Jian and Vasiloudis, Theodore and Ma, Runjie and Zhang, Houyu and Wang, Zichen and Adeshina, Soji and Nisa, Israt and others},
  booktitle={Proceedings of the 30th ACM SIGKDD Conference on Knowledge Discovery and Data Mining},
  pages={6356--6367},
  year={2024}
}

\end{document}